\documentclass[preprint,aps,12pt,showpacs,nofootinbib,tightenlines]{revtex4}
\usepackage{amsmath}
\usepackage{amssymb}
\usepackage{CJK}
\usepackage{epsfig}
\usepackage{graphicx}
\usepackage{subfigure}
\usepackage{slashed}
\begin{document}
\def\pslash{\rlap{\hspace{0.02cm}/}{p}}
\def\eslash{\rlap{\hspace{0.02cm}/}{e}}
\title {Lepton flavor violating Higgs decay $h\rightarrow \mu\tau$ in the littlest Higgs Model with
T-parity}
\author{Bingfang Yang$^{1}$}\email{yangbingfang@htu.edu.cn}
\author{Jinzhong Han$^{2}$}\email{hanjinzhong@zknu.edu.cn}
\author{Ning Liu$^{1}$}\email{wlln@mail.ustc.edu.cn}
\affiliation{ $^1$ College of Physics and Materials Science,
Henan Normal University, Xinxiang 453007, China\\
$^2$School of Physics and Telecommunications Engineering, Zhoukou
Normal University, Henan, 466001, China
   \vspace*{1.5cm}  }

\begin{abstract}

Inspired by the recent CMS $h \to \mu\tau$ excess, we calculate the
lepton flavor violating Higgs decay $h \to \mu\tau$ in the littlest
Higgs model with T-parity (LHT). Under the constraints of $\ell_i
\to \ell_j \gamma$, $Z \to \ell_i \bar{\ell}_j$ and Higgs data, we
find that the branching ratio of $h \to  \mu\tau$ can maximally
reach $\mathcal O(10^{-4})$. We also investigate the correlation
between $h \to \mu\tau $, $\tau \to \mu \gamma$ and $Z \to \mu\tau$,
which can be used to test LHT model at future $e^+e^-$ colliders.

\end{abstract}
 \maketitle
\section{ Introduction}
\noindent The discovery of the Higgs boson at the Large Hadron
Collider (LHC) \cite{LHC-higgs} is a great step toward understanding
the electroweak symmetry breaking (EWSB) mechanism. To ultimately
establish its nature, a precise study of the Higgs boson properties,
in particular the Higgs rare decays and productions \cite{exotic},
will be important tasks at LHC and future colliders.

In fact, CMS 8 TeV data has shown a 2.4¦Ò excess in searching for
Higgs mediated lepton flavor violation (LFV) process $h \to \tau
\mu$, which is interpreted to the branching ratio: $
\textrm{Br}(h\rightarrow \mu\tau) <
1.51\times10^{-2}(\textrm{CMS})$\cite{CMS}. While ATLAS only
observed a small excess in one of the signal regions and reported an
upper limit: $\textrm{Br}(h\rightarrow \mu\tau)<
1.43\times10^{-2}(\textrm{ATLAS})$\cite{ATLAS}. From the available
data it is premature to draw any definite conclusion and more data
is needed to confirm its existence. However, the lepton flavor
violating decay of the Higgs boson is widely predicted in various
extensions of the Standard Model (SM), such as seesaw\cite{seesaw},
supersymmetric (SUSY)\cite{susy}, two-Higgs doublet model
(2HDM)\cite{THD}, 3-3-1 model\cite{331}, and other
ones\cite{nonsusy}. In the SM, the LFV process is extremely
suppressed by Glashow-Iliopoulos-Maiani (GIM) mechanism\cite{GIM}
due to the smallness of neutrino mass. So, any observation of such
decays would indicate the new physics beyond the SM.

As an extension of the SM, the littlest Higgs Model with
T-parity(LHT) is one of the popular candidates that can successfully
solve the hierarchy problem. The LHT model predicts many new
particles, such as heavy gauge bosons, mirror fermions, heavy
scalars and heavy top partners. Moreover, the flavour structure of
the LHT model is richer than the one of the SM, mainly due to the
presence of the mirror fermions and their weak interactions with the
ordinary fermions. It has been shown that the LHT model can give
significant contributions to some LFV processes \cite{LHTLFV}. In
this work, we investigate the LFV process $h\rightarrow \mu\tau$ in
the LHT model under the current constraint of $\ell_i \to \ell_j
\gamma$, $Z \to \ell_i \bar{\ell}_j$ and Higgs data.

The paper is organized as follows. In Sec.II we give a brief review
of the LHT model related to our work. In Sec.III we calculate the
LFV process $h\rightarrow \mu\tau$ in unitary gauge under current
constraints. In Sec.IV we investigate the correlation between $h \to
\mu\tau $, $\tau \to \mu \gamma$ and $Z \to \mu\tau $ in the LHT
model. Finally, we draw our conclusions in Sec.V.

\section{ A brief review of the LHT model}
\noindent The LHT model is based on an $SU(5)/SO(5)$ non-linear
$\sigma$ model\cite{LHT}, where the $SU(5)$ global symmetry is
broken down to $SO(5)$ at the scale $f\sim \mathcal O$ (TeV) by the
vacuum expectation value (VEV) of the $\sigma$ field, $\Sigma_0$,
given by
\begin{eqnarray}
\Sigma_0=\langle\Sigma\rangle=
\begin{pmatrix}
{\bf 0}_{2\times2} & 0 & {\bf 1}_{2\times2} \\
                         0 & 1 &0 \\
                         {\bf 1}_{2\times2} & 0 & {\bf 0}_{2\times 2}
\end{pmatrix}.
\end{eqnarray}

The VEV $\Sigma_0$ also breaks the gauged subgroup $\left[
SU(2)\times U(1) \right]^2$ of the $SU(5)$ down to the SM
electroweak $SU(2)_L \times U(1)_Y$. After EWSB, the new T-odd gauge
bosons $W_{H}^{\pm},Z_{H},A_{H}$ acquires masses, given at $\mathcal
O(v^{2}/f^{2})$ by
\begin {equation}
M_{W_{H}}=M_{Z_{H}}=gf(1-\frac{v^{2}}{8f^{2}}),~~M_{A_{H}}=\frac{g'f}{\sqrt{5}}
(1-\frac{5v^{2}}{8f^{2}}),
\end {equation}
with $g$ and $g'$ being the SM $SU(2)$ and $U(1)$ gauge couplings,
respectively. The T-even $W^{\pm}$and $Z$ bosons of the SM, whose
masses at $\mathcal O(v^{2}/f^{2})$ are given by
\begin {equation}
M_{W}=\frac{gv}{2}(1-\frac{v^{2}}{12f^{2}}),~~M_{Z}=\frac{gv}
{2\cos\theta_{W}}(1-\frac{v^{2}}{12f^{2}}),~~M_{A}=0.
\end {equation}
Here, $v$ represents the Higgs doublet VEV, which can be given by
\begin{equation}
v = \frac{f}{\sqrt{2}} \arccos{\left( 1 -
\frac{v_\textrm{SM}^2}{f^2} \right)} \simeq v_\textrm{SM} \left( 1 +
\frac{1}{12} \frac{v_\textrm{SM}^2}{f^2} \right),
\end{equation}
where $v_{SM} = 246$ GeV is the SM Higgs VEV.

The implementation of T-parity in the fermion sector requires the
introduction of mirror fermions. Then, the T-odd mirror partners for
each SM fermions are added and one can write down a Yukawa-type
interaction to give masses to the mirror fermions
\begin{eqnarray}
\mathcal{L}_{\textrm{mirror}}=-\kappa_{ij}f\left(\bar\Psi_2^i\xi +
  \bar\Psi_1^i\Sigma_0\Omega\xi^\dagger\Omega\right)\Psi_R^j+h.c.\
\end{eqnarray}
where $i, j=1,2,3$ are the generation indices. After EWSB, the
mirror leptons acquire masses, given at $\mathcal O(v^{2}/f^{2})$ by
\begin{equation}
m_{\ell_{H}^{i}}=\sqrt{2}\kappa_if, ~~m_{\nu_{H}^{i}}=
m_{\ell_{H}^{i}}(1-\frac{v^2}{8f^2}),
\end{equation}
where $\kappa_i$ are the eigenvalues of the mass matrix $\kappa$.

As discussed in detail in Ref.\cite{F-LHT}, the existence of two
Cabibbo-Kobayashi-Maskawa (CKM)-like unitary mixing matrices
$V_{H\ell}$ and $V_{H\nu}$ is one of the important ingredients in
the mirror lepton sector. Note that $V_{H\ell}$ and $V_{H\nu}$ are
related through the Pontecorvo-Maki-Nakagata-Saki (PMNS) matrix:
\begin{equation}
V_{H\nu}^{\dag}V_{H\ell}=V^{\dag}_{\rm PMNS},
\end{equation}
where in $V_{\rm PMNS}$ the Majorana phases are set to zero as no
Majorana mass term has been introduced for right-handed neutrinos.

Follow Ref.\cite{vhd}, the matrix $V_{H\ell}$ can be parameterized
with three mixing angles
$\theta^\ell_{12},\theta^\ell_{23},\theta^\ell_{13}$ and three
complex phases $\delta^\ell_{12},\delta^\ell_{23},\delta^\ell_{13}$
\begin{eqnarray}
V_{H\ell}=
\begin{pmatrix}
c^\ell_{12}c^\ell_{13}&s^\ell_{12}c^\ell_{13}e^{-i\delta^\ell_{12}}&s^\ell_{13}e^{-i\delta^\ell_{13}}\\
-s^\ell_{12}c^\ell_{23}e^{i\delta^\ell_{12}}-c^\ell_{12}s^\ell_{23}s^\ell_{13}e^{i(\delta^\ell_{13}-\delta^\ell_{23})}&
c^\ell_{12}c^\ell_{23}-s^\ell_{12}s^\ell_{23}s^\ell_{13}e^{i(\delta^\ell_{13}-\delta^\ell_{12}-\delta^\ell_{23})}&
s^\ell_{23}c^\ell_{13}e^{-i\delta^\ell_{23}}\\
s^\ell_{12}s^\ell_{23}e^{i(\delta^\ell_{12}+\delta^\ell_{23})}-c^\ell_{12}c^\ell_{23}s^\ell_{13}e^{i\delta^\ell_{13}}&
-c^\ell_{12}s^\ell_{23}e^{i\delta^\ell_{23}}-s^\ell_{12}c^\ell_{23}s^\ell_{13}e^{i(\delta^\ell_{13}-\delta^\ell_{12})}&
c^\ell_{23}c^\ell_{13}
\end{pmatrix}
\end{eqnarray}

For the Yukawa interactions of the down-type quarks and charged
leptons, one of the possible effective Lagrangians~\cite{caseAB} is
given by
\begin{eqnarray}
{\cal L}_{\rm down} &=&\frac{i\lambda_d}{2\sqrt{2}} f \epsilon_{ij}
\epsilon_{xyz}\left[ (\bar{\Psi}'_2)_x \Sigma_{i y} \Sigma_{j z} X
-(\bar{\Psi}'_1 \Sigma_0)_x \tilde{\Sigma}_{i y} \tilde{\Sigma}_{j
z} {\tilde{X}} \right]d_{R}, \label{down-yukawa}
\end{eqnarray}
where $\Psi'_1=(-\sigma_2 q_1,0,0_2)^{\rm T}$,
$\Psi'_2=(0_2,0,-\sigma_2 q_2)^{\rm T}$, $i, j$= 1, 2 and $x, y,
z$=3, 4, 5. Here $X$ transforms into $\tilde{X}$ under T-parity, and
it is a singlet under $SU(2)_i~(i=1-2)$ and its $U(1)_i~(i=1-2)$
charges are $(Y_1,~Y_2)=(1/10,~-1/10)$. Usually, there are two
possible choices for $X$: $X=(\Sigma_{33})^{-1/4}$ (denoted as Case
A) and $X=(\Sigma^\dagger_{33})^{1/4}$ (denoted as Case B), where
$\Sigma_{33}$ is the $(3,3)$ component of the non-linear sigma model
field $\Sigma$. At order $\mathcal{O} \left( v_{SM}^4/f^4 \right)$,
the corresponding corrections to the Higgs couplings with respect to
their SM values are given by ($d \equiv d,s,b,\ell^{\pm}_i$)
\begin{eqnarray}
    \frac{g_{h \bar{d} d}}{g_{h \bar{d} d}^{SM}} &=& 1-
        \frac{1}{4} \frac{v_{SM}^{2}}{f^{2}} + \frac{7}{32}
        \frac{v_{SM}^{4}}{f^{4}} \qquad \text{Case A} \nonumber \\
    \frac{g_{h \bar{d} d}}{g_{h \bar{d} d}^{SM}} &=& 1-
        \frac{5}{4} \frac{v_{SM}^{2}}{f^{2}} - \frac{17}{32}
        \frac{v_{SM}^{4}}{f^{4}} \qquad \text{Case B}
    \label{dcoupling}
\end{eqnarray}

\section{Branching ratio for $h\rightarrow \mu\tau$ in the LHT model}

\noindent
\begin{figure}[htbp]
\scalebox{0.5}{\epsfig{file=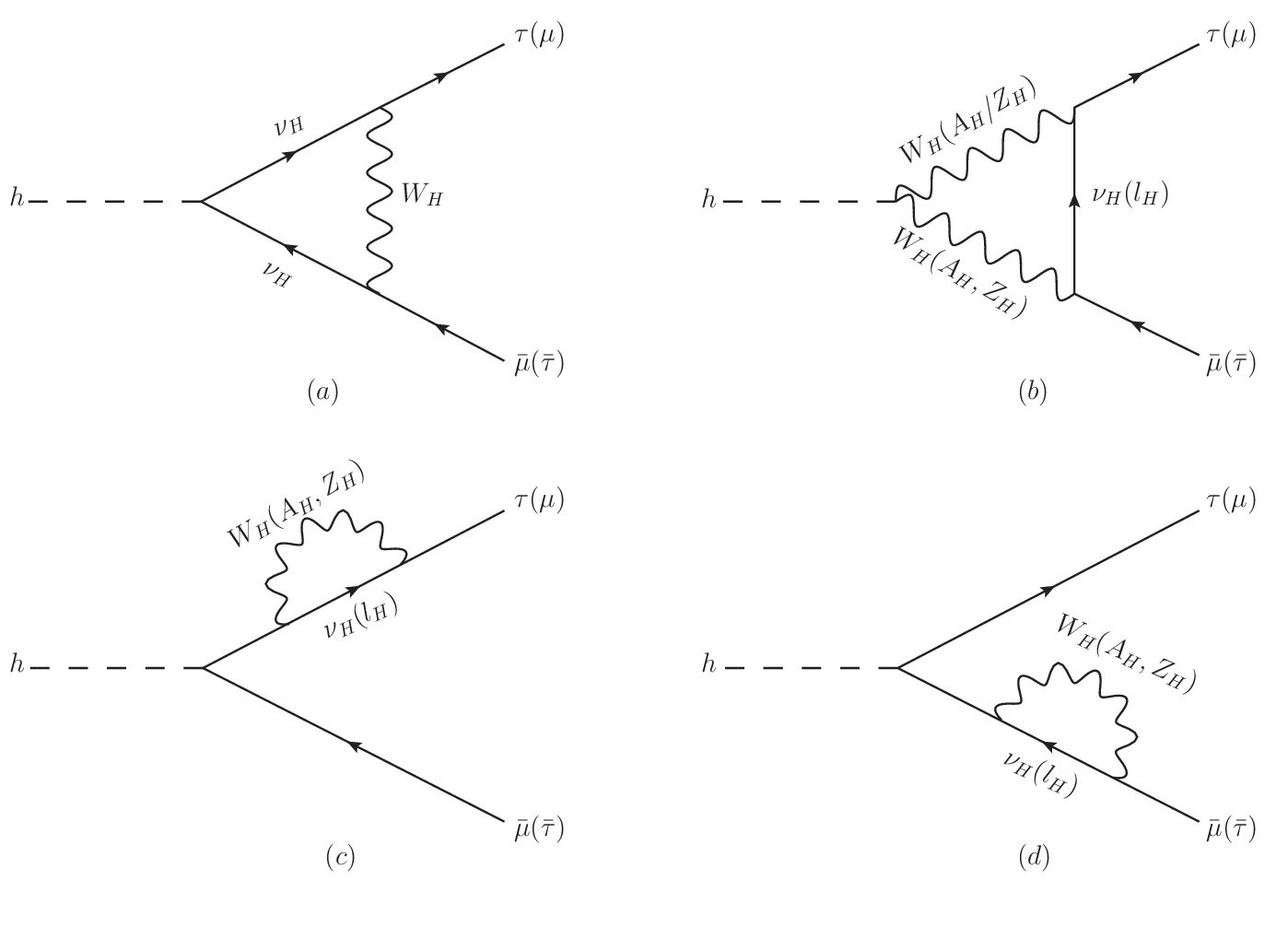}}\vspace{-0.5cm}\caption{Feynman
diagrams of the decay $h\rightarrow \mu^{\pm}\tau^{\mp}$ at one-loop
level in unitary gauge.}\label{htaumu}
\end{figure}
In the LHT model, the relevant Feynman diagrams of the decay
$h\rightarrow \mu\tau$ at one-loop level in unitary gauge are shown
in Fig.\ref{htaumu}, where the Goldstone bosons do not appear,
Fig.1(a)(b) are vertex diagrams and Fig.1(c)(d) are self-energy
diagrams. We can see that the flavor violating interactions between
SM charged leptons and mirror leptons are mediated by the heavy
gauge bosons $A_{H}, Z_{H}, W_{H}^{\pm}$. According to our
calculation, we find that the contributions of the self-energy
diagram and the contributions of the vertex diagram are at different
order, i.e. $\Gamma_{\rm vertex}\propto\mathcal{O}
(v^{2}/f^{2})\Gamma_{\rm self}$. To be clear, we show the relevant
Feynman rules and the explicit expressions of the $h\rightarrow
\mu\bar{\tau}$ invariant amplitudes in Appendix A and Appendix B,
respectively. We checked the divergence in the self-energy and vertex diagrams and found that the divergent terms have been canceled at $\mathcal{O} (v^{2}/f^{2})$ order. Since the dominant contributions come from the self-energy diagrams, we ignore the contributions of the vertex diagram in the following calculations. Each
loop diagram is composed of some scalar loop functions \cite{loop
function}, which are calculated by using LoopTools\cite{loop tools}.

In our numerical calculations, the SM parameters are taken as
follows\cite{parameters}
\begin{eqnarray}
\nonumber &&\sin^{2}\theta_{W}=0.231,~\alpha_{e}=1/128,~M_{Z}=91.1876\textrm{GeV},\\
&&m_{\mu}=105.66\textrm{MeV},~m_{\tau}=1776.82\textrm{MeV},
~m_{h}=125\textrm{GeV}.
\end{eqnarray}

The LHT parameters related to our calculations are the scale $f$,
the Yukawa couplings $\kappa_i$ of the mirror neutrinos and the
parameters in the matrices $V_{H\ell},V_{H\nu}$. For the mirror
neutrino masses, we assume
\begin{equation}
m_{\ell_{H}^{1}}=m_{\ell_{H}^{2}}=m_{\nu_{H}^{1}}=m_{\nu_{H}^{2}}=M_{12}=\sqrt{2}\kappa_{12}f,~~
m_{\ell_{H}^{3}}=m_{\nu_{H}^{3}}=M_{3}=\sqrt{2}\kappa_{3}f.
\end{equation}
For the Yukawa couplings, the search for the mono-jet events at the
LHC Run-1\cite{k-LHC1} give the constraint $\kappa_i\geq 0.6$.
Considering the constraints in Ref.\cite{constraints}, we scan over
the free parameters $f$, $\kappa_{12}$ and $\kappa_{3}$ within the
following region
\begin{eqnarray*}
500{\rm GeV}\leq f\leq 2000{\rm GeV},~~0.6\leq \kappa_{12}\leq
3,~~0.6\leq \kappa_{3}\leq 3.
\end{eqnarray*}
For the parameters in the matrices $V_{H\nu},V_{H\ell}$, we follow
Ref.\cite{case} to consider the two scenarios as follows

\begin{figure}[htbp]
\begin{center}
\scalebox{0.35}{\epsfig{file=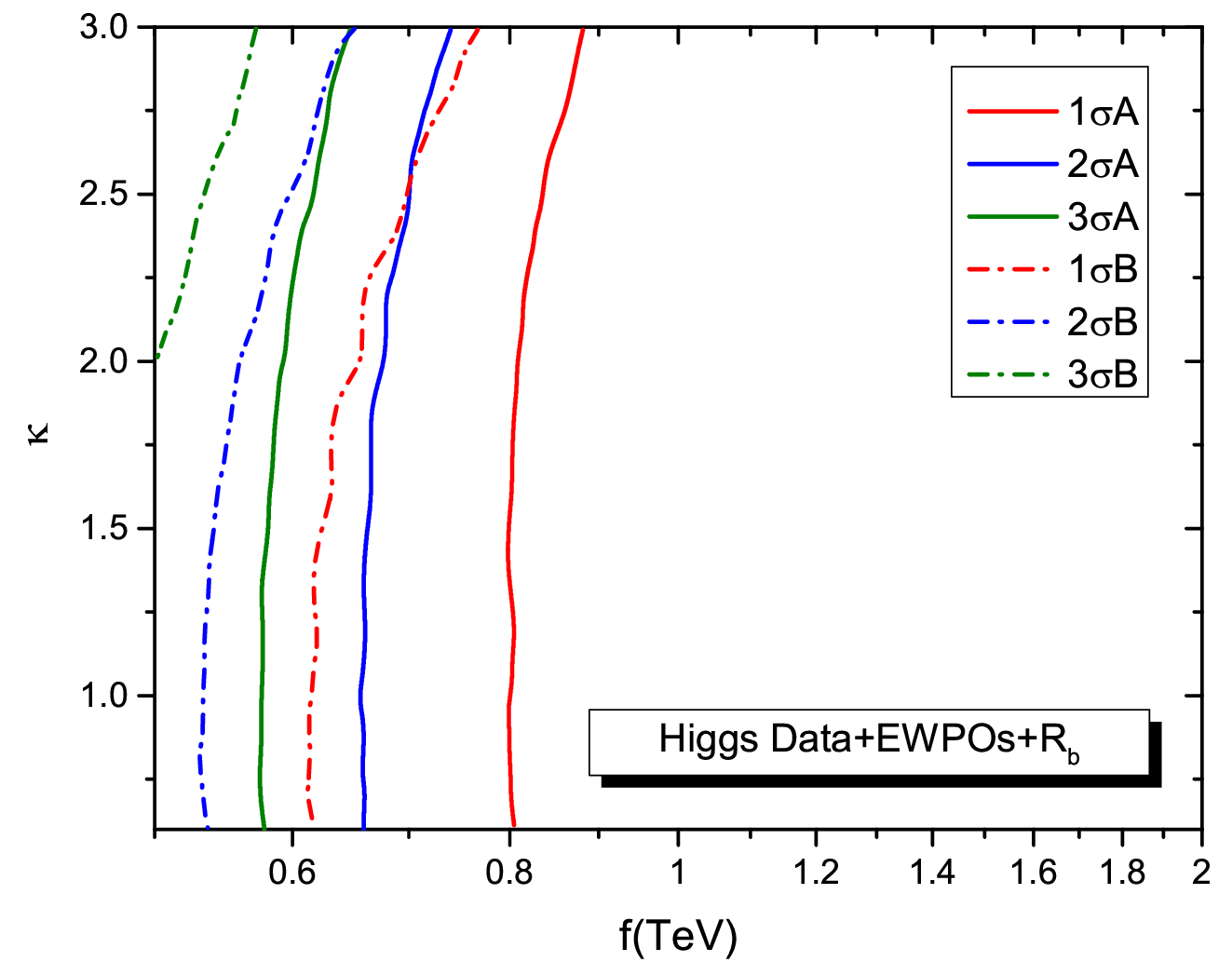}}\vspace{-0.7cm}
\caption{Excluded regions (above each contour) in the $\kappa\sim f$
plane of the LHT model for Case A and Case B, where the parameter
$R$ is marginalized over. The solid (dash) lines from right to left
respectively correspond to $1\sigma$, $2\sigma$ and $3\sigma$
exclusion limits for Case A(Case B).}\label{fkab}
\end{center}
\end{figure}

\begin{itemize}
\item {Scenario I:}
$V_{H\nu} = \textrm{I}$,$V_{H\ell} = V^{\dag}_{\rm PMNS}$;
\item {Scenario II:}
 $V_{H\ell} = V_{\rm CKM}$.
\end{itemize}
where the PMNS matrix\cite{PMNS} and CKM matrix\cite{parameters} are
given by
\begin{equation}
      V_{\rm PMNS}=\left(\begin{array}{ccc}
        0.822^{+0.010}_{-0.011} & 0.547^{+0.016}_{-0.015} & 0.155\pm0.008\\
        0.451^{+0.014}_{-0.014} & 0.648^{+0.012}_{-0.014} & 0.614^{+0.019}_{-0.017}\\
        0.347^{+0.016}_{-0.014} & 0.529^{+0.015}_{-0.014} & 0.774^{+0.013}_{-0.015}
      \end{array}\right),
\end{equation}
\begin{equation}\label{CKM}
      V_{\rm CKM}=\left(\begin{array}{ccc}
        0.97427\pm0.00014 & 0.22536\pm0.00061 & 0.00355\pm0.00015\\
        0.22522\pm0.00061 & 0.97344\pm0.00015 & 0.0414\pm0.0012\\
        0.00886^{+0.00033}_{-0.00032} & 0.0405^{+0.0011}_{-0.0012} & 0.99914\pm0.00005
      \end{array}\right).
\end{equation}
Furthermore, we will consider the constraint from the global fit of
the current Higgs data and the electroweak precision observables
(EWPOs) \cite{global fit} as shown in Fig.\ref{fkab}.

\begin{figure}[htbp]
\begin{center}
\scalebox{0.3}{\epsfig{file=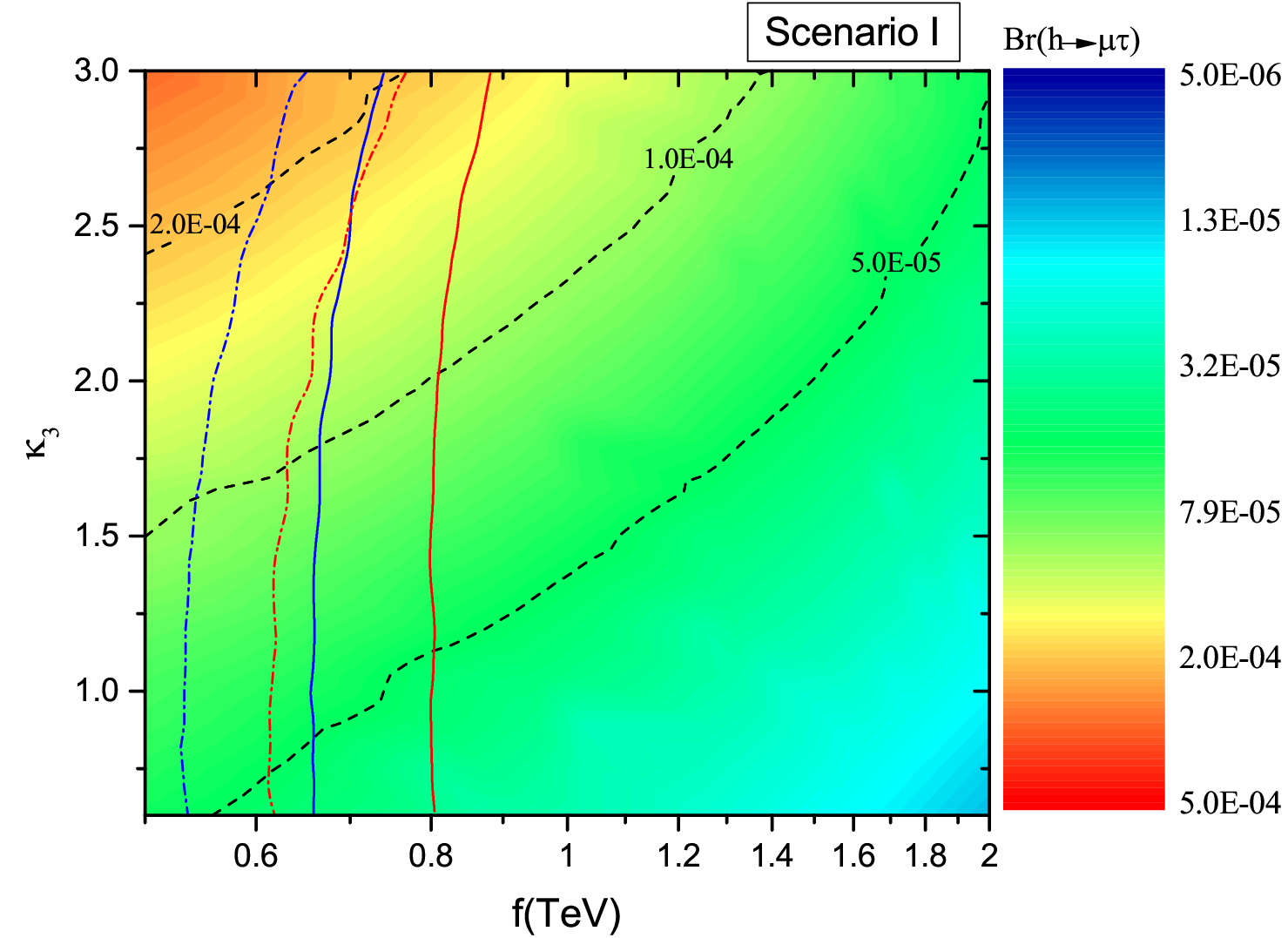}}\vspace{-0.5cm}\hspace{-0cm}
\scalebox{0.3}{\epsfig{file=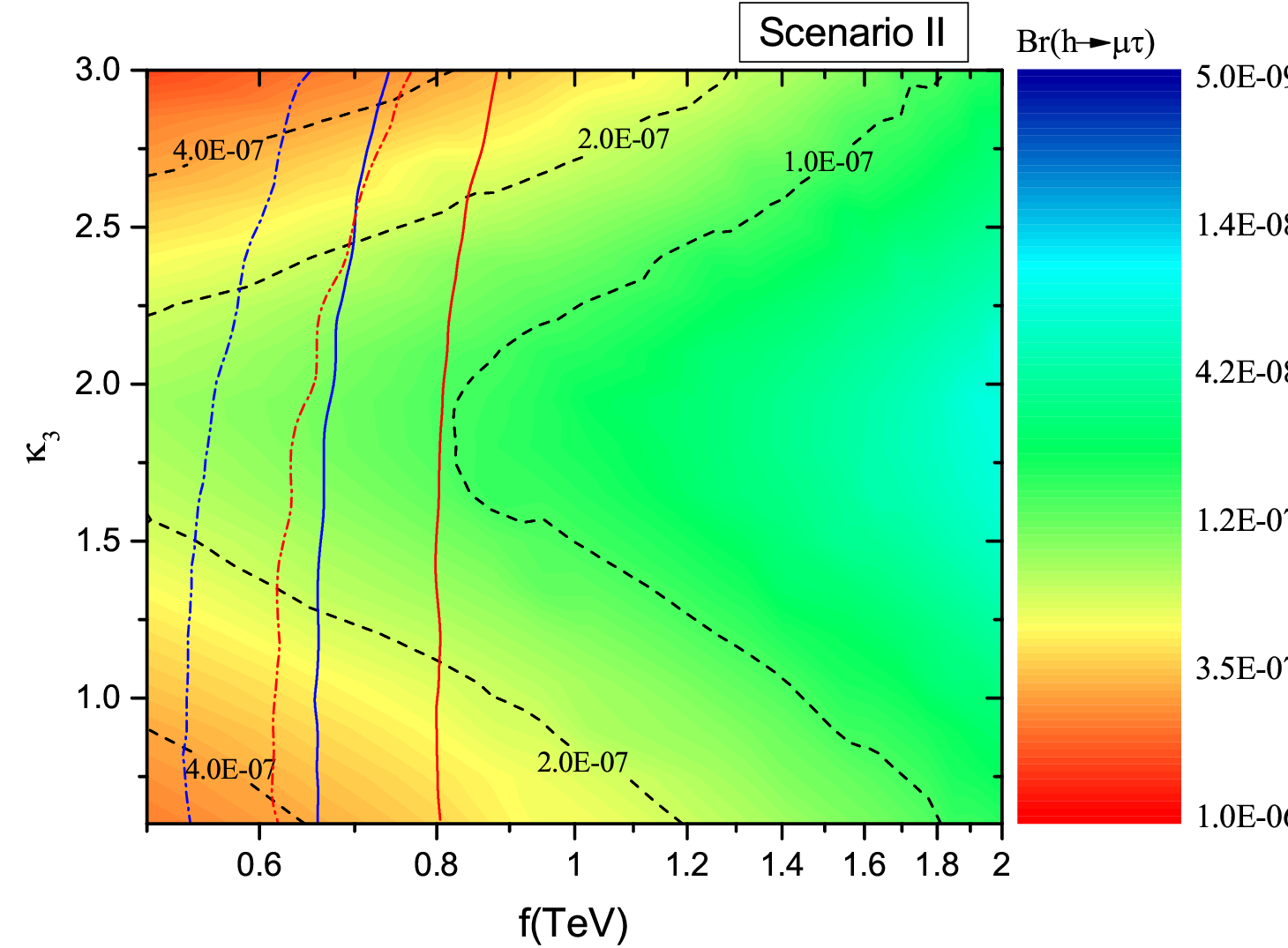}} \caption{Branching ratios of
$h\rightarrow \mu\tau$ in the $\kappa_{3}\sim f$ plane for two
scenarios with excluded regions of Case A and Case B, respectively.
The red lines and blue lines respectively correspond to $1\sigma$
and $2\sigma$ exclusion limits as shown in
Fig.\ref{fkab}.}\label{fkcase12}
\end{center}
\end{figure}
\begin{figure}[htbp]
\begin{center}
\scalebox{0.3}{\epsfig{file=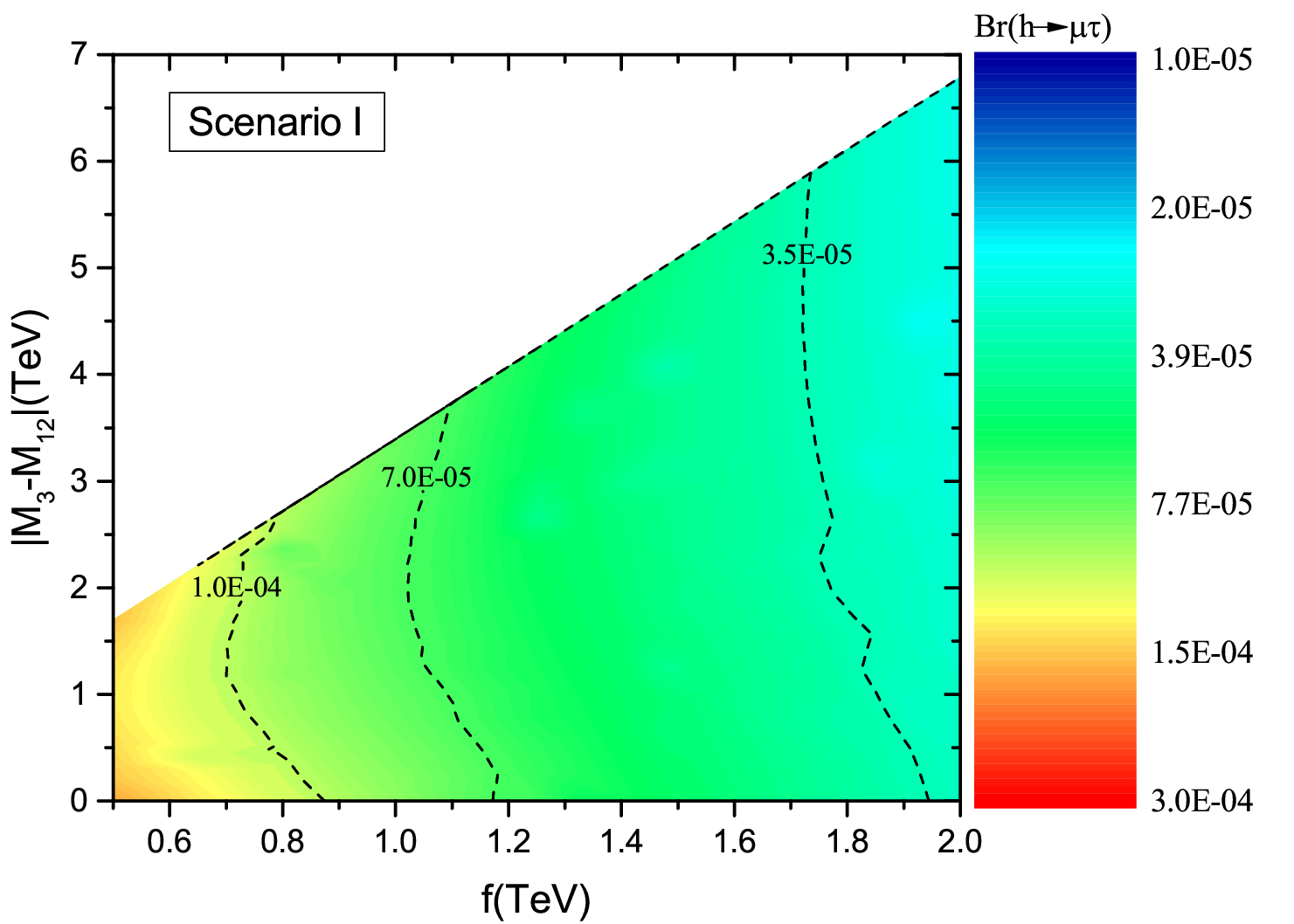}}\vspace{-0.5cm}\hspace{-0.cm}
\scalebox{0.3}{\epsfig{file=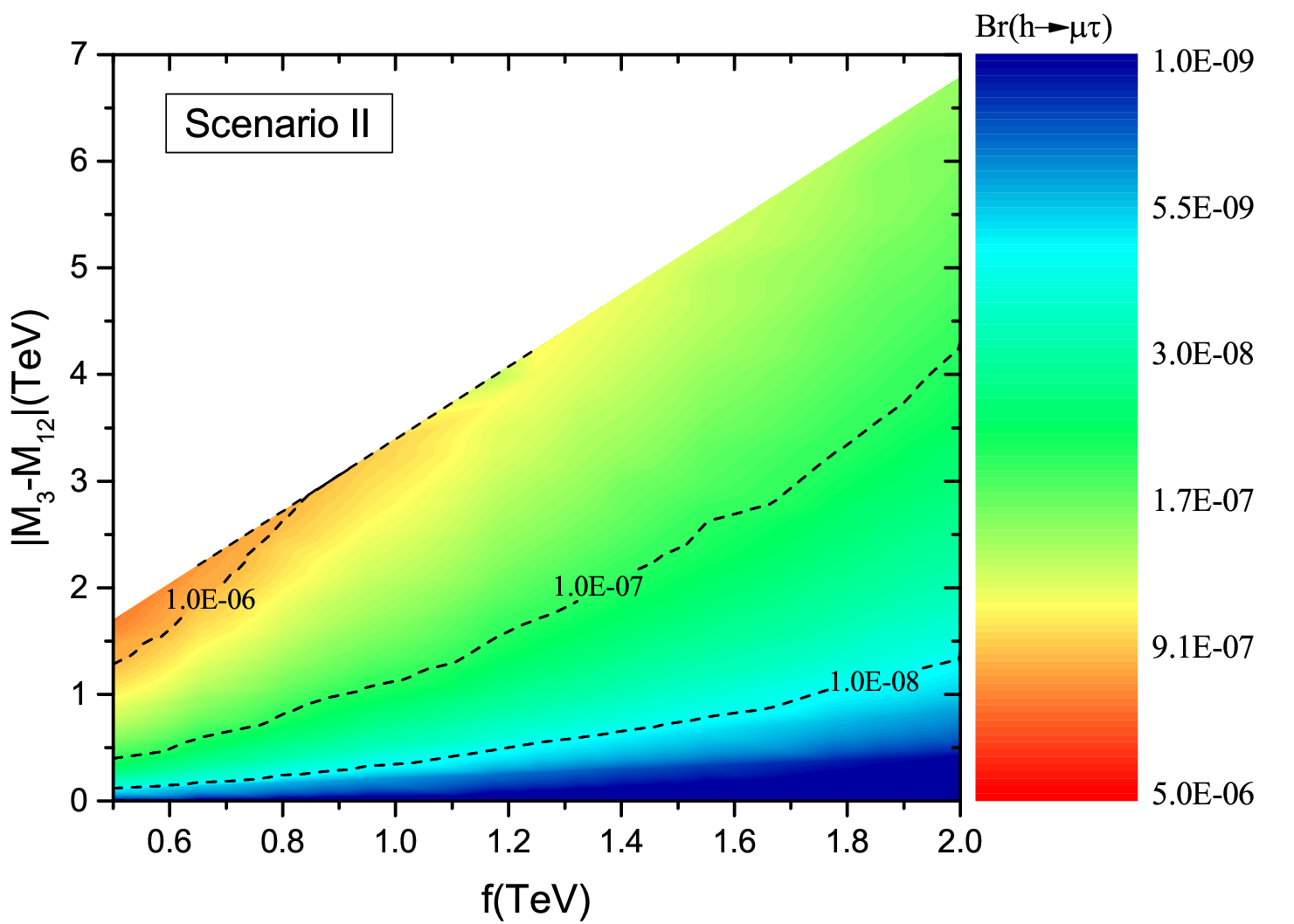}} \caption{Branching ratios of
$h\rightarrow \mu\tau$ in the $\mid M_{3}-M_{12}\mid\sim f$ plane
for two scenarios.}\label{case12}
\end{center}
\end{figure}
In Fig.\ref{fkcase12}, we show the branching ratios of $h\rightarrow
\mu\tau$ in the $\kappa_{3}\sim f$ plane for two scenarios with
excluded regions of Case A and Case B, where the $h\rightarrow
\mu\bar{\tau}$ and $h\rightarrow \tau\bar{\mu}$ modes have been
summed. From the left panel of Fig.4, we can see that the branching
ratio of $h\rightarrow \mu\tau$ in scenario I can reach about
$2\times10^{-4}$ at $2\sigma$ level for Case A, which will become
larger for Case B. From the right panel of Fig.\ref{fkcase12}, we
can see that the branching ratio of $h\rightarrow \mu\tau$ in
scenario II can reach over $4\times10^{-7}$ at $2\sigma$ level,
which is three orders of magnitude smaller than the one in scenario
I. We can see that the behaviors for two scenarios are very
different due to the different selection of the matrix $V_{H\ell}$.
From the two panels of Fig.\ref{fkcase12}, we can see that the large
branching ratios mainly lie in the upper-left corner for scenario I
and upper-left or lower-left corners for scenario II, where the
scale $f$ is small and the Yukawa coupling $\kappa_{3}$ is either
too large or too small.

In Fig.\ref{case12}, we show the branching ratios of $h\rightarrow
\mu\tau$ in the $\mid M_{3}-M_{12}\mid\sim f$ plane for two
scenarios, respectively. We can see that the branching ratio of
$h\rightarrow \mu\tau$ is insensitive to the mass splitting $\mid
M_{3}-M_{12}\mid$ values for scenario I. The largest branching
ratios lie in the region of the contour figure with small $f$ and
$\mid M_{3}-M_{12}\mid$ of $0\sim2$ TeV. For scenario II, we can see
that the branching ratio of $h\rightarrow \mu\tau$ is enhanced by
the increasing mass splitting $\mid M_{3}-M_{12}\mid$. The largest
branching ratios lie in the upper-left of the contour figure with
small $f$ and $\mid M_{3}-M_{12}\mid$ of $1\sim2$ TeV.

\section{correlation between $h\rightarrow
\mu\tau$, $\tau\rightarrow \mu\gamma$ and $Z\rightarrow \mu\tau$ }
The upper limits on the LFV processes $\tau \to \mu \gamma$ and $Z
\to \mu\tau $ are set: $ \textrm{Br}(\tau\rightarrow \mu\gamma) <
4.4\times10^{-8}$\cite{taumuga}, $ \textrm{Br}(Z\rightarrow \mu\tau)
< 1.69\times10^{-5}$\cite{zmutau}, which may further strengthen the
bounds on the branching ratios of $h\rightarrow \mu\tau$.

\begin{figure}[htbp]
\begin{center}
\scalebox{0.3}{\epsfig{file=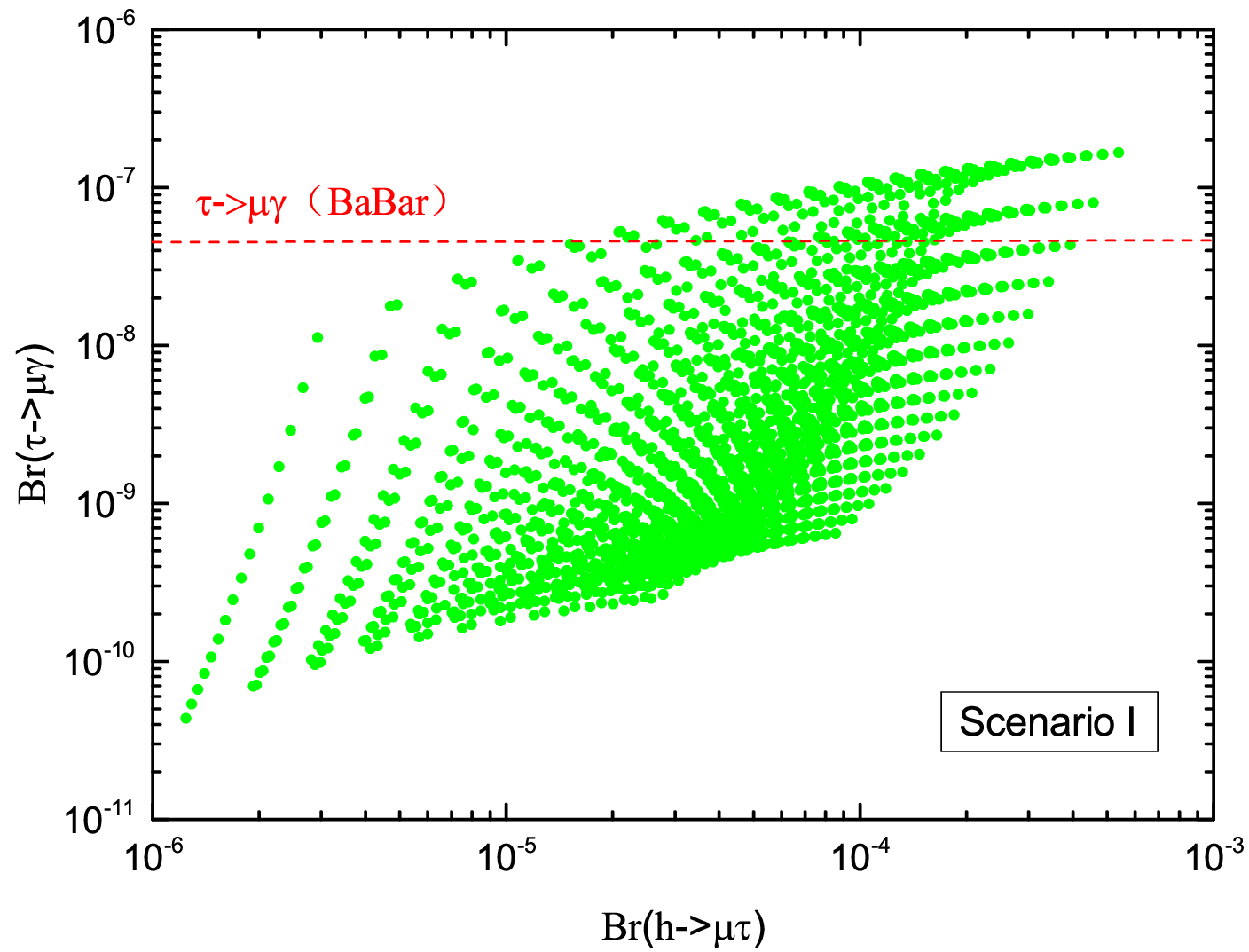}}\vspace{-0.5cm}\hspace{-0.cm}
\scalebox{0.3}{\epsfig{file=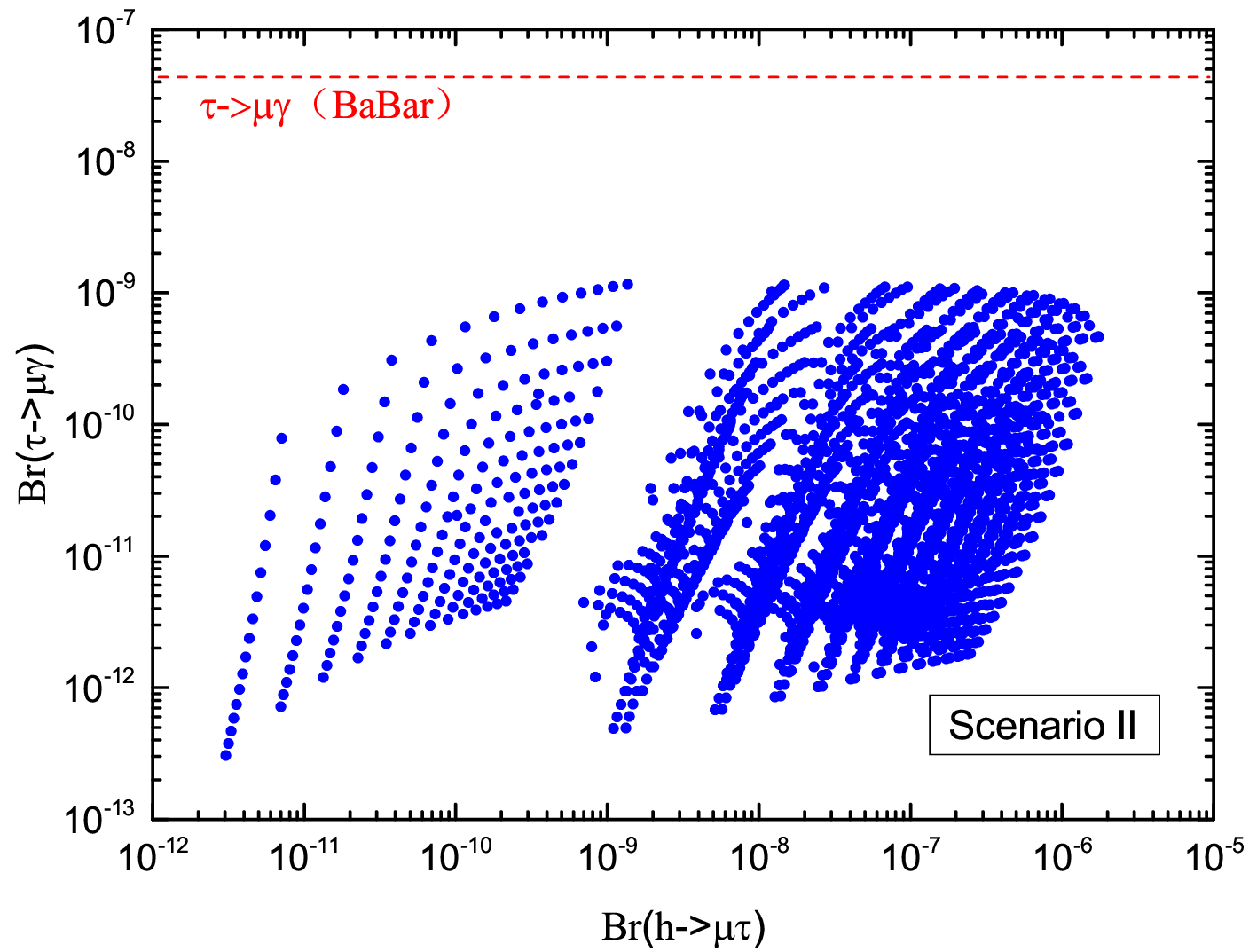}} \caption{Correlation between
Br($h\rightarrow \mu\tau$) and Br($\tau\rightarrow \mu\gamma$) for
two scenarios.}\label{htau}
\end{center}
\end{figure}
\begin{figure}[htbp]
\begin{center}
\scalebox{0.3}{\epsfig{file=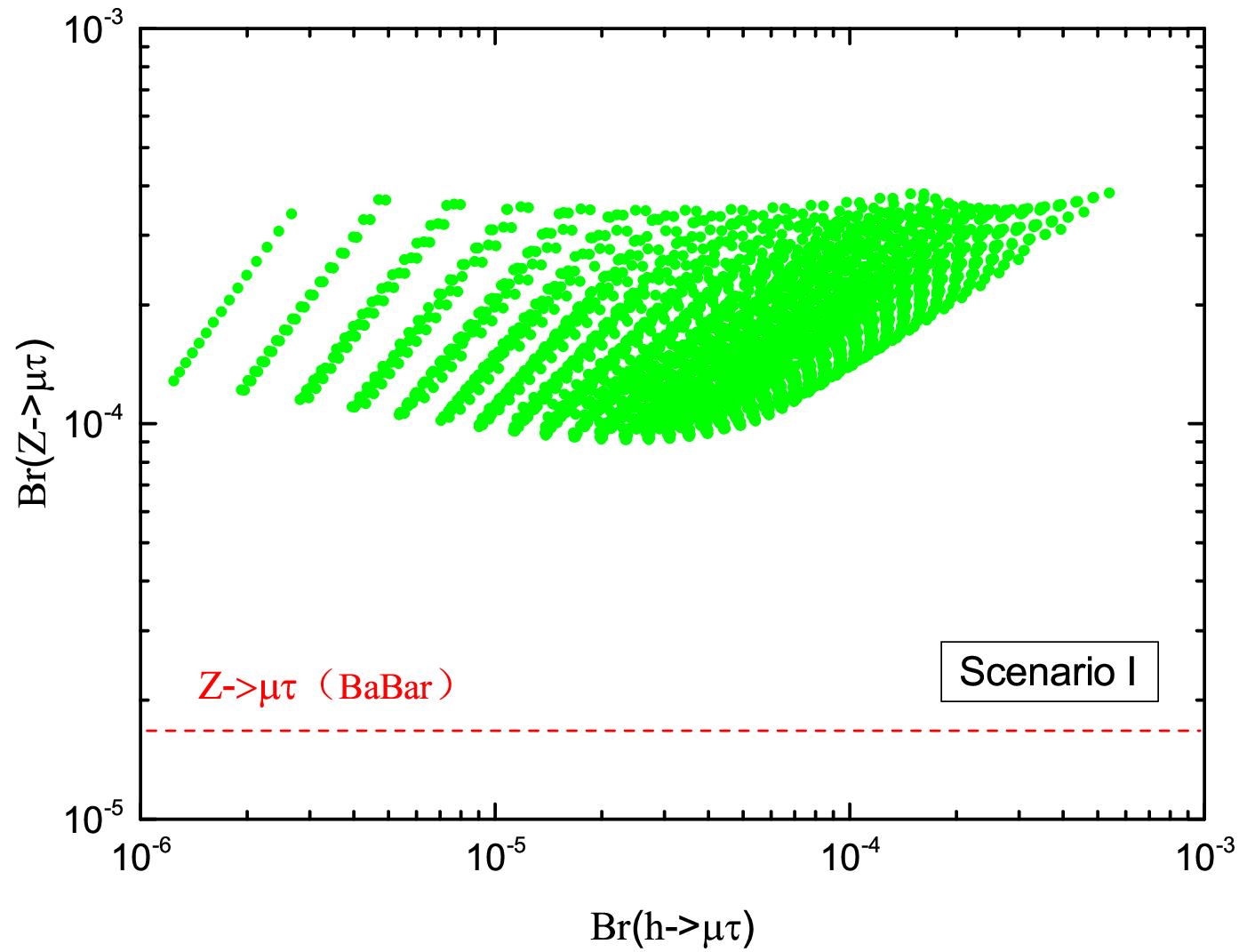}}\vspace{-0.5cm}\hspace{-0.cm}
\scalebox{0.3}{\epsfig{file=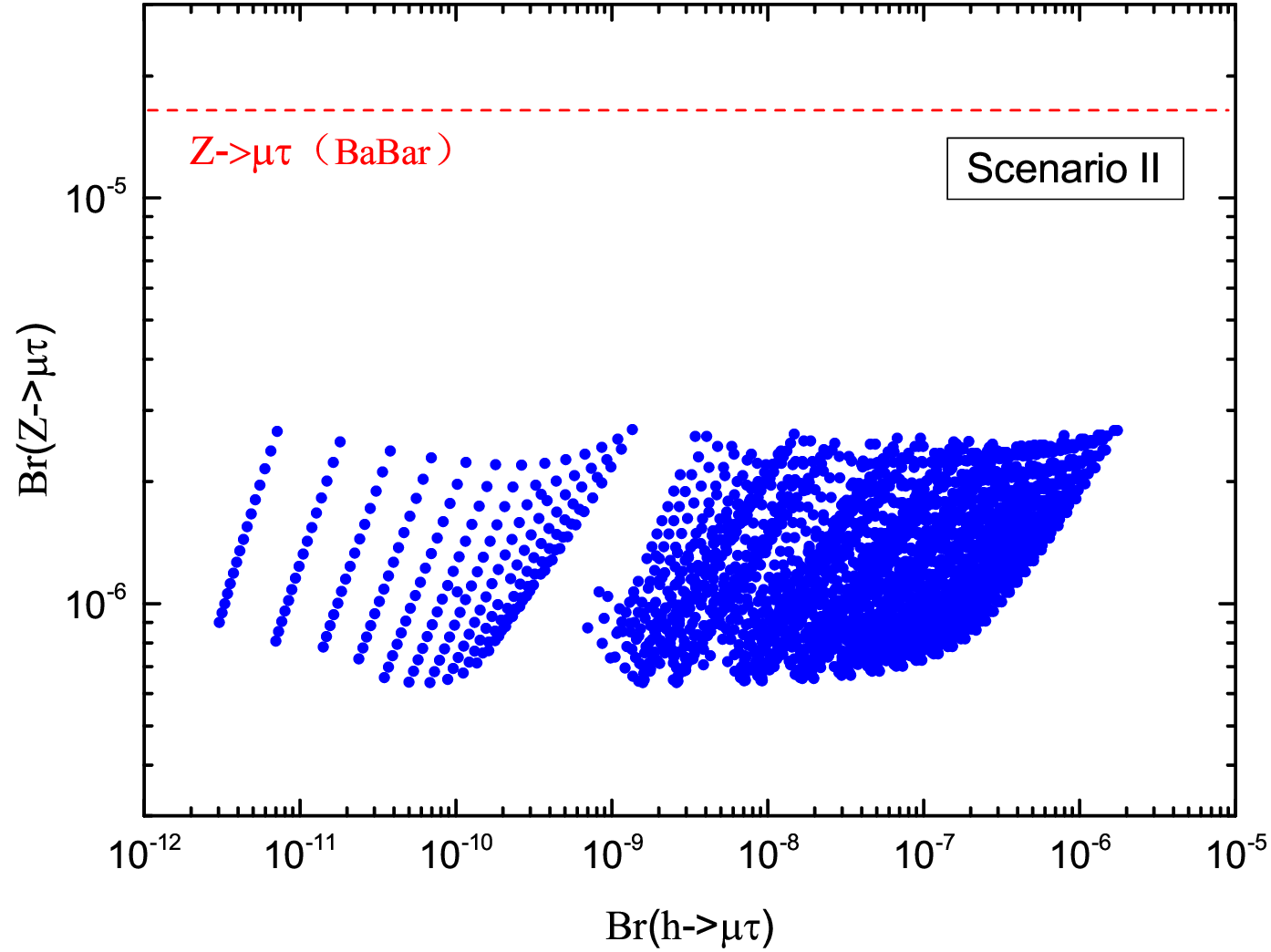}} \caption{Correlation between
Br($h\rightarrow \mu\tau$) and Br($Z\rightarrow \mu\tau$) for two
scenarios.}\label{hz}
\end{center}
\end{figure}

In Fig.\ref{htau}, we show the correlation between Br($h\rightarrow
\mu\tau$) and Br($\tau\rightarrow \mu\gamma$) in two scenarios. The
Br($\tau\rightarrow \mu\gamma$) can easily be obtained from
Eq.(3.21) in Ref.\cite{CLFV-LHT}, where we take
Br($\tau^{-}\rightarrow
\nu_{\tau}\mu^{-}\bar{\nu}_{\mu}$)=$(17.41\pm 0.04)\%$, and few
other studies on such processes can be found in
Ref.\cite{Tau-muon-LHT}. In Scenario I, we can see that a minority
of points is outside the allowed range, which implies that the
$V_{H\ell}$ matrix must be more hierarchical than
$V_{\textrm{PMNS}}$ in order to satisfy the present upper bounds on
$h\rightarrow \mu\tau$ and $\tau\rightarrow \mu\gamma$. In Scenario
II, we can see that all the points are in allowed range, this is
because $V_{\textrm{CKM}}$ matrix is much more hierarchical than
$V_{\textrm{PMNS}}$.

In Fig.\ref{hz}, we show the correlation between Br($h\rightarrow
\mu\tau$) and Br($Z\rightarrow \mu\tau$) in two scenarios. The
partial $Z$ decay width $\Gamma(Z\rightarrow\mu\tau)$ can be
calculated by using LoopTools, the relevant Feynman diagrams can be
found in Refs.\cite{LFV-LHT}. In Scenario I, we can see that all the
points violate the current experimental bounds and the great
majority of points exceed $\mathcal O(10^{-4})$. This will require
that the $V_{H\ell}$ matrix must be more hierarchical than
$V_{\textrm{PMNS}}$, unless the mirror lepton masses are
quasi-degenerate. In Scenario II, we can see that all the points
satisfy the current upper bounds due to the large hierarchy of the
$V_{\textrm{CKM}}$ matrix.

\section{Conclusions}

\noindent

In this paper, we calculated LFV Higgs decay $h\rightarrow \mu\tau$
at one-loop level in the LHT model. According to the parameters in
the mixing matrices, we considered two scenarios and found that the
branching ratios for $h\rightarrow \mu\tau$ can respectively reach
$\mathcal O(10^{-4})$ and $\mathcal O(10^{-7})$ under the current
experimental constraints. We also investigated the correlation
between $h \to \mu\tau $, $\tau \to \mu \gamma$ and $Z \to \mu\tau$,
and found that the $Z \to \mu\tau $ can give a substantial
constraint on the $h \to \mu\tau $, which required that the
$V_{H\ell}$ matrix must have a very different hierarchy from
$V_{\textrm{PMNS}}$ matrix.

\section*{Acknowledgement}
We thank Lei Wu, Qinghong Cao and Junjie Cao for helpful suggestions
and discussions. This work is supported by the National Natural
Science Foundation of China (NNSFC) under grant Nos. 11405047,
11305049, by Specialized Research Fund for the Doctoral Program of
Higher Education under Grant No. 20134104120002, by the Startup
Foundation for Doctors of Henan Normal University under Grant Nos.
11112 and qd15207, and by the Education Department Foundation of
Henan Province(14A140010).

\vspace{0.5cm}

\textbf{Appendix A: Feynman rules}

\begin{table}[h]
\begin{center}
{\small
\begin{tabular}{c|c|c|c}
\hline \hline
 {  \, Interaction\,\, } & { \, Feynman rule\,\, }& {  \, Interaction\,\, }& { \, Feynman rule\,\, } \\[0.04cm]
 \hline
 $H\bar{\nu}_{H}^{j}\nu_{H}^{j}$ & $i\frac{m_{\nu^{j}_{H}}}{v}\frac{v^{2}}{4f^{2}}$&$HA_{H\alpha}A_{H\beta}$ & $-\frac{i}{2}g'^{2}vg_{\alpha\beta}$           \\[0.02cm]
 $\bar{\ell}_{H}^{j}A_{H\alpha}\ell^{k}$ & $-\frac{ig'}{10}(V_{H\ell})_{jk}\gamma_{\alpha}P_{L}$&$HZ_{H\alpha}Z_{H\beta}$ & $-\frac{i}{2}g^{2}vg_{\alpha\beta}$           \\[0.02cm]
 $\bar{\ell}_{H}^{j}Z_{H\alpha}\ell^{k}$ & $\frac{ig}{2}(V_{H\ell})_{jk}\gamma_{\alpha}P_{L}$&$HA_{H\alpha}Z_{H\beta}$ & $-\frac{i}{2}g'gvg_{\alpha\beta}$          \\[0.02cm]
 $\bar{\nu}_{H}^{j}W_{H\alpha}^{+}\ell^{k}$ & $-\frac{ig}{\sqrt{2}}(V_{H\ell})_{jk}\gamma_{\alpha}P_{L}$&$HW_{H\alpha}^{+}W_{H\beta}^{-}$ & $-\frac{i}{2}g^{2}vg_{\alpha\beta}$           \\[0.02cm]
 \hline \hline
\end{tabular}
} \\[0.1cm]

\end{center}\label{tab:htaumu}
\end{table}

\textbf{Appendix B: The expression of the $h\rightarrow
\mu\bar{\tau}$ invariant amplitudes}

They can be represented in form of 1-point, 2-point and 3-point
standard functions $A,B_{0},B_{1},C_{ij}$. Here the momenta
$p_{\mu}, p_{\tau}$ are assumed to be outgoing.

(1)\textbf{The vertex diagram contribution:}
\begin{eqnarray}
&&\Gamma^{W_{H}\nu_{H}}_{(a)}=-\frac{g^{2}}{2}(V_{H\ell})_{j2}(V_{H\ell})^\ast_{j3}\frac{m^{2}_{\nu_{H}^{j}}v}{4f^{2}}\frac{i}{16\pi^{2}}\nonumber\\
&&[4C_{\beta}\gamma^{\beta}+2C_{0}(\pslash _{\mu}-\pslash
_{\tau})+\frac{2}{m^{2}_{W_{H}}}\tilde{C}_{\beta}\gamma^{\beta}
+\frac{1}{m^{2}_{W_{H}}}C_{\beta\alpha}\gamma^{\beta}(\pslash
_{\mu}-\pslash
_{\tau})\gamma^{\alpha}]P_{L}\\
&&C_{\beta}=C_{\beta}(p_{\mu},-p_{h},m_{W_{H}},m_{\nu_{H}^{j}},m_{\nu_{H}^{j}}),\nonumber\\
&&\tilde{C}_{\beta}=p_{\mu\beta}[-B_{0}(-p_{h},m_{\nu_{H}^{j}},m_{\nu_{H}^{j}})+m^{2}_{W_{H}}C_{11}]-p_{h\beta}[B_{1}+m^{2}_{W_{H}}C_{12}]\nonumber
\end{eqnarray}
\begin{eqnarray}
&&\Gamma^{W_{H}W_{H}}_{(b)}=\frac{g^{2}}{2}\frac{g^{2}v}{2}(V_{H\ell})_{j2}(V_{H\ell})^\ast_{j3}\frac{i}{16\pi^{2}}\nonumber\\
&&[2C_{\beta}\gamma^{\beta}+\frac{2}{m^{2}_{W_{H}}}\tilde{C}_{\beta}\gamma^{\beta}-\frac{m^{4}_{\nu_{H}^{j}}}{m^{4}_{W_{H}}}C_{\beta}\gamma^{\beta}
-\frac{\gamma^{\beta}(2p_{\mu}-p_{h})^{\alpha}\tilde{C}_{\beta\alpha}+(2\pslash_{\mu}-\pslash_{h})\tilde{C'}_{0}}{m^{4}_{W_{H}}}\nonumber\\&&-\frac{1}{m^{4}_{W_{H}}}(\tilde{B}_{\beta}\gamma^{\beta}-\gamma^{\beta}\pslash
_{\mu}\gamma^{\alpha}B_{\beta\alpha}+m^{2}_{\nu_{H}^{j}}B_{\beta}\gamma^{\beta}-m^{2}_{\nu_{H}^{j}}\pslash
_{\mu}B_{0})]P_{L}\\
&&B_{\beta}=B_{\beta}(-p_{h},m_{W_{H}},m_{W_{H}}),
\tilde{B}_{\beta}=-p_{h\beta}[m^{2}_{W_{H}}B_{1}-A_{0}(m_{W_{H}})],\nonumber\\
&&B_{\beta\alpha}=p_{\mu\beta}p_{\tau\alpha}B_{21}+g_{\beta\alpha}B_{22}(-p_{h},m_{W_{H}},m_{W_{H}}),\nonumber\\
&&C_{\beta}=C_{\beta}(p_{\mu},-p_{h},m_{\nu_{H}^{j}},m_{W_{H}},m_{W_{H}}),\nonumber\\
&&\tilde{C}_{\beta}=p_{\mu\beta}[-B_{0}(-p_{h},m_{W_{H}},m_{W_{H}})+m^{2}_{\nu_{H}^{j}}C_{11}]-p_{h\beta}[B_{1}+m^{2}_{\nu_{H}^{j}}C_{12}]\nonumber\\
&&\tilde{C'}_{0}=g^{\alpha\beta}B_{\alpha\beta}(-p_{h},m_{W_{H}},m_{W_{H}})+2p_{\mu}p_{h}B_{1}+p_{\mu}^{2}B_{0}+m^{2}_{\nu_{H}^{j}}(B_{0}+m^{2}_{\nu_{H}^{j}}C_{0})\nonumber\\
&&\tilde{C}_{\beta\alpha}=p_{\mu\beta}p_{\mu\alpha}(B_{0}+m^{2}_{W_{H}}C_{21})+p_{h\beta}p_{h\alpha}(B_{21}+m^{2}_{W_{H}}C_{22})\nonumber\\&&~~~~~-(p_{\mu\beta}p_{h\alpha}+p_{\mu\alpha}p_{h\beta})(-B_{1}+m^{2}_{W_{H}}C_{23})
+g_{\beta\alpha}(B_{22}+m^{2}_{W_{H}}C_{24})
\end{eqnarray}
\begin{eqnarray}
&&\Gamma^{A_{H}A_{H}}_{(b)}=\frac{g'^{2}}{100}\frac{g'^{2}v}{2}(V_{H\ell})_{j2}(V_{H\ell})^\ast_{j3}\frac{i}{16\pi^{2}}\nonumber\\
&&[2C_{\beta}\gamma^{\beta}+\frac{2}{m^{2}_{A_{H}}}\tilde{C}_{\beta}\gamma^{\beta}-\frac{m^{4}_{\ell_{H}^{j}}}{m^{4}_{A_{H}}}C_{\beta}\gamma^{\beta}
-\frac{\gamma^{\beta}(2p_{\mu}-p_{h})^{\alpha}\tilde{C}_{\beta\alpha}+(2\pslash_{\mu}-\pslash_{h})\tilde{C'}_{0}}{m^{4}_{A_{H}}}\nonumber\\&&-\frac{1}{m^{4}_{A_{H}}}(\tilde{B}_{\beta}\gamma^{\beta}-\gamma^{\beta}\pslash
_{\mu}\gamma^{\alpha}B_{\beta\alpha}+m^{2}_{l_{H}^{j}}B_{\beta}\gamma^{\beta}-m^{2}_{l_{H}^{j}}\pslash
_{\mu}B_{0})]P_{L}\\
&&B_{\beta}=B_{\beta}(-p_{h},m_{A_{H}},m_{A_{H}}),
\tilde{B}_{\beta}=-p_{h\beta}[m^{2}_{A_{H}}B_{1}-A_{0}(m_{A_{H}})],\nonumber\\
&&B_{\beta\alpha}=p_{\mu\beta}p_{\tau\alpha}B_{21}+g_{\beta\alpha}B_{22}(-p_{h},m_{A_{H}},m_{A_{H}}),\nonumber\\
&&C_{\beta}=C_{\beta}(p_{\mu},-p_{h},m_{\ell_{H}^{j}},m_{A_{H}},m_{A_{H}}),\nonumber\\
&&\tilde{C}_{\beta}=p_{\mu\beta}[-B_{0}(-p_{h},m_{A_{H}},m_{A_{H}})+m^{2}_{\ell_{H}^{j}}C_{11}]-p_{h\beta}[B_{1}+m^{2}_{\ell_{H}^{j}}C_{12}]\nonumber\\
&&\tilde{C'}_{0}=g^{\alpha\beta}B_{\alpha\beta}(-p_{h},m_{A_{H}},m_{A_{H}})+2p_{\mu}p_{h}B_{1}+p_{\mu}^{2}B_{0}+m^{2}_{l_{H}^{j}}(B_{0}+m^{2}_{l_{H}^{j}}C_{0})\nonumber\\
&&\tilde{C}_{\beta\alpha}=p_{\mu\beta}p_{\mu\alpha}(B_{0}+m^{2}_{A_{H}}C_{21})+p_{h\beta}p_{h\alpha}(B_{21}+m^{2}_{A_{H}}C_{22})\nonumber\\&&~~~~~-(p_{\mu\beta}p_{h\alpha}+p_{\mu\alpha}p_{h\beta})(-B_{1}+m^{2}_{A_{H}}C_{23})
+g_{\beta\alpha}(B_{22}+m^{2}_{A_{H}}C_{24})
\end{eqnarray}
\begin{eqnarray}
&&\Gamma^{A_{H}Z_{H}}_{(b)}=\frac{g'^{2}}{20}\frac{g^{2}v}{2}(V_{H\ell})_{j2}(V_{H\ell})^\ast_{j3}\frac{i}{16\pi^{2}}\nonumber\\
&&[-2C_{\beta}\gamma^{\beta}-\frac{1}{m^{2}_{A_{H}}}\tilde{C}_{\beta}\gamma^{\beta}-\frac{1}{m^{2}_{Z_{H}}}\tilde{C}_{\beta}\gamma^{\beta}+\frac{m^{4}_{\ell_{H}^{j}}}{m^{2}_{A_{H}}m^{2}_{Z_{H}}}C_{\beta}\gamma^{\beta}\nonumber\\
&&+\frac{(\tilde{B}_{\beta}\gamma^{\beta}-\gamma^{\beta}\pslash
_{\mu}\gamma^{\alpha}B_{\beta\alpha}+m^{2}_{l_{H}^{j}}B_{\beta}\gamma^{\beta}-m^{2}_{l_{H}^{j}}\pslash
_{\mu}B_{0})}{m^{2}_{A_{H}}m^{2}_{Z_{H}}}\nonumber\\
&&+\frac{\gamma^{\beta}(2p_{\mu}-p_{h})^{\alpha}\tilde{C}_{\beta\alpha}+(2\pslash_{\mu}-\pslash_{h})\tilde{C'}_{0}}{m^{2}_{A_{H}}m^{2}_{Z_{H}}}]P_{L}\\
&&B_{\beta}=B_{\beta}(-p_{h},m_{A_{H}},m_{Z_{H}}),
\tilde{B}_{\beta}=-p_{h\beta}[m^{2}_{A_{H}}B_{1}-A_{0}(m_{Z_{H}})],\nonumber\\
&&B_{\beta\alpha}=p_{\mu\beta}p_{\tau\alpha}B_{21}+g_{\beta\alpha}B_{22}(-p_{h},m_{A_{H}},m_{Z_{H}}),\nonumber\\
&&C_{\beta}=C_{\beta}(p_{\mu},-p_{h},m_{\ell_{H}^{j}},m_{A_{H}},m_{Z_{H}}),\nonumber\\
&&\tilde{C}_{\beta}=p_{\mu\beta}[-B_{0}(-p_{h},m_{A_{H}},m_{Z_{H}})+m^{2}_{\ell_{H}^{j}}C_{11}]-p_{h\beta}[B_{1}+m^{2}_{\ell_{H}^{j}}C_{12}]\nonumber\\
&&\tilde{C'}_{0}=g^{\alpha\beta}B_{\alpha\beta}(-p_{h},m_{A_{H}},m_{Z_{H}})+2p_{\mu}p_{h}B_{1}+p_{\mu}^{2}B_{0}+m^{2}_{l_{H}^{j}}(B_{0}+m^{2}_{l_{H}^{j}}C_{0})\nonumber\\
&&\tilde{C}_{\beta\alpha}=p_{\mu\beta}p_{\mu\alpha}(B_{0}+m^{2}_{A_{H}}C_{21})+p_{h\beta}p_{h\alpha}(B_{21}+m^{2}_{A_{H}}C_{22})\nonumber\\&&~~~~~-(p_{\mu\beta}p_{h\alpha}+p_{\mu\alpha}p_{h\beta})(-B_{1}+m^{2}_{A_{H}}C_{23})
+g_{\beta\alpha}(B_{22}+m^{2}_{A_{H}}C_{24})
\end{eqnarray}
\begin{eqnarray}
&&\Gamma^{Z_{H}A_{H}}_{(b)}=\frac{g'^{2}}{20}\frac{g^{2}v}{2}(V_{H\ell})_{j2}(V_{H\ell})^\ast_{j3}\frac{i}{16\pi^{2}}\nonumber\\
&&[-2C_{\beta}\gamma^{\beta}-\frac{1}{m^{2}_{A_{H}}}\tilde{C}_{\beta}\gamma^{\beta}-\frac{1}{m^{2}_{Z_{H}}}\tilde{C}_{\beta}\gamma^{\beta}+\frac{m^{4}_{\ell_{H}^{j}}}{m^{2}_{A_{H}}m^{2}_{Z_{H}}}C_{\beta}\gamma^{\beta}\nonumber\\
&&+\frac{(\tilde{B}_{\beta}\gamma^{\beta}-\gamma^{\beta}\pslash
_{\mu}\gamma^{\alpha}B_{\beta\alpha}+m^{2}_{l_{H}^{j}}B_{\beta}\gamma^{\beta}-m^{2}_{l_{H}^{j}}\pslash
_{\mu}B_{0})}{m^{2}_{A_{H}}m^{2}_{Z_{H}}}\nonumber\\
&&+\frac{\gamma^{\beta}(2p_{\mu}-p_{h})^{\alpha}\tilde{C}_{\beta\alpha}+(2\pslash_{\mu}-\pslash_{h})\tilde{C'}_{0}}{m^{2}_{A_{H}}m^{2}_{Z_{H}}}]P_{L}\\
&&B_{\beta}=B_{\beta}(-p_{h},m_{Z_{H}},m_{A_{H}}),
\tilde{B}_{\beta}=-p_{h\beta}[m^{2}_{Z_{H}}B_{1}-A_{0}(m_{A_{H}})],\nonumber\\
&&B_{\beta\alpha}=p_{\mu\beta}p_{\tau\alpha}B_{21}+g_{\beta\alpha}B_{22}(-p_{h},m_{Z_{H}},m_{A_{H}}),\nonumber\\
&&C_{\beta}=C_{\beta}(p_{\mu},-p_{h},m_{\ell_{H}^{j}},m_{Z_{H}},m_{A_{H}}),\nonumber\\
&&\tilde{C}_{\beta}=p_{\mu\beta}[-B_{0}(-p_{h},m_{Z_{H}},m_{A_{H}})+m^{2}_{\ell_{H}^{j}}C_{11}]-p_{h\beta}[B_{1}+m^{2}_{\ell_{H}^{j}}C_{12}]\nonumber\\
&&\tilde{C'}_{0}=g^{\alpha\beta}B_{\alpha\beta}(-p_{h},m_{Z_{H}},m_{A_{H}})+2p_{\mu}p_{h}B_{1}+p_{\mu}^{2}B_{0}+m^{2}_{l_{H}^{j}}(B_{0}+m^{2}_{l_{H}^{j}}C_{0})\nonumber\\
&&\tilde{C}_{\beta\alpha}=p_{\mu\beta}p_{\mu\alpha}(B_{0}+m^{2}_{Z_{H}}C_{21})+p_{h\beta}p_{h\alpha}(B_{21}+m^{2}_{Z_{H}}C_{22})\nonumber\\&&~~~~~-(p_{\mu\beta}p_{h\alpha}+p_{\mu\alpha}p_{h\beta})(-B_{1}+m^{2}_{Z_{H}}C_{23})
+g_{\beta\alpha}(B_{22}+m^{2}_{Z_{H}}C_{24})
\end{eqnarray}
\begin{eqnarray}
&&\Gamma^{Z_{H}Z_{H}}_{(b)}=\frac{g^{2}}{4}\frac{g^{2}v}{2}(V_{H\ell})_{j2}(V_{H\ell})^\ast_{j3}\frac{i}{16\pi^{2}}\nonumber\\
&&[2C_{\beta}\gamma^{\beta}+\frac{2}{m^{2}_{Z_{H}}}\tilde{C}_{\beta}\gamma^{\beta}-\frac{m^{4}_{\ell_{H}^{j}}}{m^{4}_{Z_{H}}}C_{\beta}\gamma^{\beta}
-\frac{\gamma^{\beta}(2p_{\mu}-p_{h})^{\alpha}\tilde{C}_{\beta\alpha}+(2\pslash_{\mu}-\pslash_{h})\tilde{C'}_{0}}{m^{4}_{Z_{H}}}\nonumber\\&&-\frac{1}{m^{4}_{Z_{H}}}(\tilde{B}_{\beta}\gamma^{\beta}-\gamma^{\beta}\pslash
_{\mu}\gamma^{\alpha}B_{\beta\alpha}+m^{2}_{l_{H}^{j}}B_{\beta}\gamma^{\beta}-m^{2}_{l_{H}^{j}}\pslash
_{\mu}B_{0})]P_{L}\\
&&B_{\beta}=B_{\beta}(-p_{h},m_{Z_{H}},m_{Z_{H}}),
\tilde{B}_{\beta}=-p_{h\beta}[m^{2}_{Z_{H}}B_{1}-A_{0}(m_{Z_{H}})],\nonumber\\
&&B_{\beta\alpha}=p_{\mu\beta}p_{\tau\alpha}B_{21}+g_{\beta\alpha}B_{22}(-p_{h},m_{Z_{H}},m_{Z_{H}}),\nonumber\\
&&C_{\beta}=C_{\beta}(p_{\mu},-p_{h},m_{\ell_{H}^{j}},m_{Z_{H}},m_{Z_{H}}),\nonumber\\
&&\tilde{C}_{\beta}=p_{\mu\beta}[-B_{0}(-p_{h},m_{Z_{H}},m_{Z_{H}})+m^{2}_{\ell_{H}^{j}}C_{11}]-p_{h\beta}[B_{1}+m^{2}_{\ell_{H}^{j}}C_{12}]\nonumber\\
&&\tilde{C'}_{0}=g^{\alpha\beta}B_{\alpha\beta}(-p_{h},m_{Z_{H}},m_{Z_{H}})+2p_{\mu}p_{h}B_{1}+p_{\mu}^{2}B_{0}+m^{2}_{l_{H}^{j}}(B_{0}+m^{2}_{l_{H}^{j}}C_{0})\nonumber\\
&&\tilde{C}_{\beta\alpha}=p_{\mu\beta}p_{\mu\alpha}(B_{0}+m^{2}_{Z_{H}}C_{21})+p_{h\beta}p_{h\alpha}(B_{21}+m^{2}_{Z_{H}}C_{22})\nonumber\\&&~~~~~-(p_{\mu\beta}p_{h\alpha}+p_{\mu\alpha}p_{h\beta})(-B_{1}+m^{2}_{Z_{H}}C_{23})
+g_{\beta\alpha}(B_{22}+m^{2}_{Z_{H}}C_{24})
\end{eqnarray}

(2)\textbf{The self-energy diagram contribution:}
\begin{eqnarray}
&&\Gamma^{\mu A_{H}}_{(c)}=\frac{g'^{2}}{100}\frac{m_{\mu}}{v}(V_{H\ell})_{j2}(V_{H\ell})^\ast_{j3}\frac{1}{p_{\tau}^{2}-m_{\mu}^{2}}\frac{i}{16\pi^{2}}\nonumber\\
&&[2(\pslash
_{\tau}+m_{\mu})\pslash
_{\tau}P_{L}B_{0}-2(\pslash
_{\tau}+m_{\mu})B_{\beta}\gamma^{\beta}P_{L}-\frac{{1}}{m^{2}_{A_{H}}}(\pslash
_{\tau}+m_{\mu})\tilde{B}_{\beta}\gamma^{\beta}P_{L}\nonumber\\
&&+\frac{(\pslash
	_{\tau}+m_{\mu})\gamma^{\beta}\pslash
	_{\tau}\gamma^{\alpha}B_{\beta\alpha}P_{L}}{m^{2}_{A_{H}}}]\nonumber\\
&&B_{\beta}=B_{\beta}(p_{\tau},m_{\ell_{H}^{j}},m_{A_{H}}),
\tilde{B}_{\beta}=p_{\tau\beta}[m^{2}_{\ell_{H}^{j}}B_{1}-A_{0}(m_{A_{H}})]\nonumber\\
&&B_{\beta\alpha}=p_{\mu\beta}p_{\tau\alpha}B_{21}+g_{\beta\alpha}B_{22}(p_{\tau},m_{\ell_{H}^{j}},m_{A_{H}})
\end{eqnarray}
\begin{eqnarray}
&&\Gamma^{\mu Z_{H}}_{(c)}=\frac{g^{2}}{4}\frac{m_{\mu}}{v}(V_{H\ell})_{j2}(V_{H\ell})^\ast_{j3}\frac{1}{p_{\tau}^{2}-m_{\mu}^{2}}\frac{i}{16\pi^{2}}\nonumber\\
&&[2(\pslash
_{\tau}+m_{\mu})\pslash
_{\tau}P_{L}B_{0}-2(\pslash
_{\tau}+m_{\mu})B_{\beta}\gamma^{\beta}P_{L}-\frac{{1}}{m^{2}_{Z_{H}}}(\pslash
_{\tau}+m_{\mu})\tilde{B}_{\beta}\gamma^{\beta}P_{L}\nonumber\\
&&+\frac{(\pslash
	_{\tau}+m_{\mu})\gamma^{\beta}\pslash
	_{\tau}\gamma^{\alpha}B_{\beta\alpha}P_{L}}{m^{2}_{Z_{H}}}]\nonumber\\
&&B_{\beta}=B_{\beta}(p_{\tau},m_{\ell_{H}^{j}},m_{Z_{H}}),
\tilde{B}_{\beta}=p_{\tau\beta}[m^{2}_{\ell_{H}^{j}}B_{1}-A_{0}(m_{Z_{H}})]\nonumber\\
&&B_{\beta\alpha}=p_{\mu\beta}p_{\tau\alpha}B_{21}+g_{\beta\alpha}B_{22}(p_{\tau},m_{\ell_{H}^{j}},m_{Z_{H}})
\end{eqnarray}
\begin{eqnarray}
&&\Gamma^{\mu W_{H}}_{(c)}=\frac{g^{2}}{2}\frac{m_{\mu}}{v}(V_{H\ell})_{j2}(V_{H\ell})^\ast_{j3}\frac{1}{p_{\tau}^{2}-m_{\mu}^{2}}\frac{i}{16\pi^{2}}\nonumber\\
&&[2(\pslash
_{\tau}+m_{\mu})\pslash
_{\tau}P_{L}B_{0}-2(\pslash
_{\tau}+m_{\mu})B_{\beta}\gamma^{\beta}P_{L}-\frac{{1}}{m^{2}_{W_{H}}}(\pslash
_{\tau}+m_{\mu})\tilde{B}_{\beta}\gamma^{\beta}P_{L}\nonumber\\
&&+\frac{(\pslash
	_{\tau}+m_{\mu})\gamma^{\beta}\pslash
	_{\tau}\gamma^{\alpha}B_{\beta\alpha}P_{L}}{m^{2}_{W_{H}}}]\nonumber\\
&&B_{\beta}=B_{\beta}(p_{\tau},m_{\nu_{H}^{j}},m_{W_{H}}),
\tilde{B}_{\beta}=p_{\tau\beta}[m^{2}_{\nu_{H}^{j}}B_{1}-A_{0}(m_{W_{H}})]\nonumber\\
&&B_{\beta\alpha}=p_{\mu\beta}p_{\tau\alpha}B_{21}+g_{\beta\alpha}B_{22}(p_{\tau},m_{\nu_{H}^{j}},m_{W_{H}})
\end{eqnarray}
\begin{eqnarray}
&&\Gamma^{\tau A_{H}}_{(d)}=\frac{g'^{2}}{100}\frac{m_{\tau}}{v}(V_{H\ell})_{j2}(V_{H\ell})^\ast_{j3}\frac{1}{p_{\mu}^{2}-m_{\tau}^{2}}\frac{i}{16\pi^{2}}\nonumber\\
&&[2\pslash
_{\mu}P_{L}(\pslash
_{\mu}+m_{\tau})B_{0}-2B_{\beta}\gamma^{\beta}P_{L}(\pslash
_{\mu}+m_{\tau})-\frac{{1}}{m^{2}_{A_{H}}}\tilde{B}_{\beta}\gamma^{\beta}P_{L}(\pslash
_{\mu}+m_{\tau})\nonumber\\
&&+\frac{\gamma^{\beta}\pslash
_{\mu}\gamma^{\alpha}B_{\beta\alpha}P_{L}(\pslash
_{\mu}+m_{\tau})}{m^{2}_{A_{H}}}]\nonumber\\
&&B_{\beta}=B_{\beta}(-p_{\mu},m_{\ell_{H}^{j}},m_{A_{H}}),
\tilde{B}_{\beta}=-p_{\mu\beta}[m^{2}_{\ell_{H}^{j}}B_{1}-A_{0}(m_{A_{H}})]\nonumber\\
&&B_{\beta\alpha}=p_{\mu\beta}p_{\tau\alpha}B_{21}+g_{\beta\alpha}B_{22}(-p_{\mu},m_{\ell_{H}^{j}},m_{A_{H}})
\end{eqnarray}
\begin{eqnarray}
&&\Gamma^{\tau Z_{H}}_{(d)}=\frac{g^{2}}{4}\frac{m_{\tau}}{v}(V_{H\ell})_{j2}(V_{H\ell})^\ast_{j3}\frac{1}{p_{\mu}^{2}-m_{\tau}^{2}}\frac{i}{16\pi^{2}}\nonumber\\
&&[2\pslash
_{\mu}P_{L}(\pslash
_{\mu}+m_{\tau})B_{0}-2B_{\beta}\gamma^{\beta}P_{L}(\pslash
_{\mu}+m_{\tau})-\frac{{1}}{m^{2}_{Z_{H}}}\tilde{B}_{\beta}\gamma^{\beta}P_{L}(\pslash
_{\mu}+m_{\tau})\nonumber\\
&&+\frac{\gamma^{\beta}\pslash
	_{\mu}\gamma^{\alpha}B_{\beta\alpha}P_{L}(\pslash
	_{\mu}+m_{\tau})}{m^{2}_{Z_{H}}}]\nonumber\\
&&B_{\beta}=B_{\beta}(-p_{\mu},m_{\ell_{H}^{j}},m_{Z_{H}}),
\tilde{B}_{\beta}=-p_{\mu\beta}[m^{2}_{\ell_{H}^{j}}B_{1}-A_{0}(m_{Z_{H}})]\nonumber\\
&&B_{\beta\alpha}=p_{\mu\beta}p_{\tau\alpha}B_{21}+g_{\beta\alpha}B_{22}(-p_{\mu},m_{\ell_{H}^{j}},m_{Z_{H}})
\end{eqnarray}
\begin{eqnarray}
&&\Gamma^{\tau W_{H}}_{(d)}=\frac{g^{2}}{2}\frac{m_{\tau}}{v}(V_{H\ell})_{j2}(V_{H\ell})^\ast_{j3}\frac{1}{p_{\mu}^{2}-m_{\tau}^{2}}\frac{i}{16\pi^{2}}\nonumber\\
&&[2\pslash
_{\mu}P_{L}(\pslash
_{\mu}+m_{\tau})B_{0}-2B_{\beta}\gamma^{\beta}P_{L}(\pslash
_{\mu}+m_{\tau})-\frac{{1}}{m^{2}_{W_{H}}}\tilde{B}_{\beta}\gamma^{\beta}P_{L}(\pslash
_{\mu}+m_{\tau})\nonumber\\
&&+\frac{\gamma^{\beta}\pslash
	_{\mu}\gamma^{\alpha}B_{\beta\alpha}P_{L}(\pslash
	_{\mu}+m_{\tau})}{m^{2}_{W_{H}}}]\nonumber\\
&&B_{\beta}=B_{\beta}(-p_{\mu},m_{\nu_{H}^{j}},m_{W_{H}}),
\tilde{B}_{\beta}=-p_{\mu\beta}[m^{2}_{\nu_{H}^{j}}B_{1}-A_{0}(m_{W_{H}})]\nonumber\\
&&B_{\beta\alpha}=p_{\mu\beta}p_{\tau\alpha}B_{21}+g_{\beta\alpha}B_{22}(-p_{\mu},m_{\nu_{H}^{j}},m_{W_{H}})
\end{eqnarray}


\begin{thebibliography}\\

\bibitem{LHC-higgs}G. Aad et al.(ATLAS Collaboration), Phys. Lett. B 710, 49 (2012);
S. Chatrachyan et al.(CMS Collaboration), Phys. Lett. B 710, 26
(2012).

\bibitem{exotic} see examples,
  A.~Falkowski and R.~Vega-Morales,
  JHEP 1412, 037 (2014),arXiv:1405.1095;
  D.~Curtin {\it et al.},
  Phys.\ Rev.\ D  90, no. 7, 075004 (2014),
  arXiv:1312.4992;
  C.~Han, N.~Liu, L.~Wu, J.~M.~Yang and Y.~Zhang,
  Eur.\ Phys.\ J.\ C 73, no. 12, 2664 (2013),
  arXiv:1212.6728;
  J.~Cao, L.~Wu, P.~Wu and J.~M.~Yang,
  JHEP 1309, 043 (2013),
  arXiv:1301.4641;
  J.~Huang, T.~Liu, L.~T.~Wang and F.~Yu,
  Phys.\ Rev.\ D 90, no. 11, 115006 (2014),
  arXiv:1407.0038;
  J.~Cao, C.~Han, L.~Wu, J.~M.~Yang and M.~Zhang,
  Eur.\ Phys.\ J.\ C 74, no. 9, 3058 (2014),
  arXiv:1404.1241;
  C.~Han, X.~Ji, L.~Wu, P.~Wu and J.~M.~Yang,
  JHEP 1404, 003 (2014),
  arXiv:1307.3790;
  S.~L.~Hu, N.~Liu, J.~Ren and L.~Wu,
  J.\ Phys.\ G 41, no. 12, 125004 (2014),
  arXiv:1402.3050;
  L.~Wu,
  JHEP 1502, 061 (2015),
  arXiv:1407.6113;
  D.~Curtin and C.~B.~Verhaaren,
  JHEP 1512, 072 (2015),
  arXiv:1506.06141;
  L.~Wu, J.~M.~Yang, C.~P.~Yuan and M.~Zhang,
  Phys.\ Lett.\ B 747, 378 (2015)
  arXiv:1504.06932;
  A.~Kobakhidze, L.~Wu and J.~Yue,
  JHEP 1410, 100 (2014),
  arXiv:1406.1961;
  JHEP 1604, 011 (2016),
  arXiv:1512.08922;
  H.~B¨¦lusca-Ma\"{\i}to and A.~Falkowski,
  arXiv:1602.02645.

\bibitem{CMS}V. Khachatryan et al. (CMS Collaboration), Phys. Lett. B 749, 337 (2015), arXiv:1502.07400.

\bibitem{ATLAS}G. Aad et al. (ATLAS Collaboration), JHEP 1511, 211 (2015), arXiv:1508.03372.


\bibitem{seesaw}E. Arganda, M. J. Herrero, X. Marcano and C. Weiland, Phys. Rev. D 91, 015001
(2015); E. Arganda, A.M. Curiel, M.J. Herrero, D. Temes, Phys. Rev.
D 71, 035011 (2005), hep-ph/0407302.

\bibitem{susy}
  J.~Cao, L.~Wu and J.~M.~Yang,
  Nucl.\ Phys.\ B 829, 370 (2010);
  E. Arganda, M. J. Herrero, X. Marcano, C. Weiland, Phys. Rev. D 93, 055010 (2016),
arXiv: 1508.04623; E. Arganda, M. J. Herrero, R. Morales and A.
Szynkman, JHEP 1603, 055 (2016), arxiv:1510.04685; A. Hammad, S.
Khalil and C. Un, arXiv:1605.07567; M. Arana-Catania, E. Arganda,
M.J. Herrero JHEP 1309, 160 (2013), arXiv:1304.3371.

\bibitem{THD}A. Crivellin, G. D¡¯Ambrosio, J. Heeck, Phys. Rev. Lett. 114, 151801 (2015); N. Bizot, S.
Davidson, M. Frigerio, J. L. Kneur, JHEP 1603, 073 (2016); N. Bizot,
S. Davidson, M. Frigerio, J. -L. Kneur, JHEP 1603, 073 (2016); F. J.
Botella, G. C. Branco, M. Nebot, M. N. Rebelo, Eur. Phys. J. C 76,
161 (2016); D. Das and A. Kundu, Phys. Rev. D 92, 015009 (2015),
arXiv:1504.01125; D. Aristizabal Sierra, A. Vicente, Phys. Rev. D
90, 115004 (2014), arXiv:1409.7690.

\bibitem{331}L. T. Hue, H. N. Long, T. T. Thuc and T. Phong Nguyen, Nucl. Phys. B 907, 37 (2016),
arXiv:1512.03266.

\bibitem{nonsusy}S. Baek, Z.-F. Kang, JHEP 1603, 106 (2016); S. Baek, K. Nishiwaki , Phys. Rev. D 93,
015002 (2016); K. Cheung, W. Y. Keung, P. Y. Tseng, Phys. Rev. D 93,
015010 (2016); W. Altmannshofer, S. Gori, A. L. Kagan, L.
Silvestrini, J. Zupan, Phys. Rev. D 93, 031301 (2016); X. G. He, J.
Tandean, Y. J. Zheng, JHEP 1509, 093 (2015); I. Dorsner, S. Fajfer,
A. Greljo, J. F. Kamenik, N. Kosnik, Ivan Nisandzic, JHEP 1506, 108
(2015); A. Crivellin, G. DAmbrosio and J. Heeck, Phys. Rev. D 91,
075006 (2015); L. D . Lima, C. S. Machado, R. D. Matheus, L. A. F.
D. Prado, JHEP 1511, 074 (2015); I. d. M. Varzielas, O. Fischer, V.
Maurer, JHEP 1508, 080 (2015); C. F. Chang, C. H. V. Chang, C. S.
Nugroho, T. C. Yuan, arXiv:1602.00680; C. H. Chen, T. Nomura,
arXiv:1602.07519; K. Huitu, V. Keus, N. Koivunen, O. Lebedev,
arXiv:1603.06614; M. Sher, K. Thrasher, Phys. Rev. D 93, 055021
(2016); A. Lami, P. Roig, Phys. Rev. D 94, 056001 (2016),
arXiv:1603.09663; J. G. Koerner, A. Pilaftsis, K. Schilcher, Phys.
Rev. D 47:1080-1086 (1993); A. Pilaftsis, Phys. Lett. B 285,
68-74(1992); J. Heeck, M. Holthausen, W. Rodejohann, Y. Shimizu,
Nucl. Phys. B 896, 281-310(2015); J. Herrero-Garcia, N. Rius, A.
Santamaria, JHEP 1611, 084(2016).

\bibitem{GIM}
S. L. Glashow, J. Iliopoulos and L. Maiani, Phys. Rev. D 2, 1285
(1970).

\bibitem{LHTLFV}A. Goyal, hep-ph/0609095; S.R. Choudhury, A.S. Cornell, A. Deandrea, N. Gaur
and A. Goyal, Phys. Rev. D 75, 055011 (2007), hep-ph/0612327.

\bibitem{LHT}H. C. Cheng and I. Low, JHEP 0309, 051 (2003); JHEP 0408, 061 (2004);
I. Low, JHEP 0410, 067 (2004); J. Hubisz and P. Meade, Phys. Rev. D
71, 035016 (2005).

\bibitem{F-LHT}J. Hubisz, S.J. Lee and G. Paz, JHEP 06,
041 (2006), hep-ph/0512169.

\bibitem{vhd}M. Blanke, et al., Phys. Lett. B 646, 253 (2007).

\bibitem{caseAB} C. R. Chen, K. Tobe and C. P. Yuan, Phys. Lett. B 640, 263 (2006).


\bibitem{loop function}G. 't Hooft and M. J. G. Veltman,  Nucl. Phys. B 153, 365 (1979).


\bibitem{loop tools}T. Hahn and M. Perez-Victoria, Comput. Phys. Commun. 118, 153
(1999); T. Hahn, Nucl. Phys. Proc. Suppl. 135, 333 (2004).

\bibitem{parameters}K. A. Olive et al., (Particle Data Group), Chinese Physics C Vol. 38, No. 9 (2014) 090001.

\bibitem{constraints} B. F. Yang, X. L. Wang and
J. Z. Han, Nucl. Phys. B 847, 1 (2011), arXiv:1103.2506; J. Hubisz,
P. Meade, A. Noble and M. Perelstein, JHEP 0601, 135 (2006); A.
Belyaev, C. R. Chen, K. Tobe and C. P. Yuan, Phys. Rev. D 74, 115020
(2006); Q. H. Cao and C. R. Chen, Phys. Rev. D 76, 075007 (2007); J.
Reuter, M. Tonini, JHEP 0213,077 (2013); X. F. Han, L. Wang, J. M.
Yang, J. Y. Zhu, Phys. Rev. D 87, 055004 (2013); J. Reuter, M.
Tonini, M. de Vries, arXiv:1307.5010; C. C.~Han, A.~Kobakhidze,
N.~Liu, L.~Wu and B. F.~Yang, Nucl.\ Phys.\ B 890, 388 (2014),
arXiv:1405.1498.


\bibitem{k-LHC1}J. Hubisz, P. Meade, A. Noble and M. Perelstein, JHEP 01, 135
(2006); M. Perelstein, J. Shao, Phys. Lett. B 704, 510 (2011); G.
Cacciapaglia, et al., Phys. Rev. D 87, 075006 (2013); J. Reuter, M.
Tonini, M. de Vries, JHEP 1402, 053 (2014); J. Reuter, M. Tonini,
JHEP 1302, 077 (2013).

\bibitem{case}H. S. Hou, Phys. Rev. D 75, 094010 (2007);
J. Z. Han, B. Z. Li and X. L. Wang, Phys. Rev. D 83, 034032 (2011),
arXiv:1102.4402.

\bibitem{PMNS}Y. J. Zhang, X. Y. Zhang, B.-Q. Ma, Phys. Rev. D 86,
093019 (2012).

\bibitem{global fit}B. F. Yang, G. F. Mi, N. Liu, JHEP 1410,
47 (2014), arXiv:1407.6123;  N.~Liu, L.~Wu, B. F.~Yang and
M.C.~Zhang, Phys.\ Lett.\ B 753, 664 (2016), arXiv:1508.07116; B. F.
Yang, Z. Y. Liu, N. Liu, arXiv:1603.04242.

\bibitem{taumuga}K. Hayasaka et al. (Belle collaboration), Phys. Lett. B 666, 16-22 (2008); B. Aubert et al. (BABAR collaboration), Phys. Rev. Lett. 104,
021802 (2010).

\bibitem{zmutau}ATLAS Collaboration, CERN-EP-2016-055, arXiv:1604.07730
[hep-ex].

\bibitem{CLFV-LHT}M. Blanke, et al., JHEP 0705, 013 (2007).

\bibitem{Tau-muon-LHT}T. Goto, Y. Okada, Y. Yamamoto, Phys. Rev. D
83, 053011 (2011).

\bibitem{LFV-LHT}J. Z. Han, X. L. Wang, B. F. Yang, Nucl. Phys. B 843, 383-395 (2011), arXiv:1101.3598; C.-X. Yue, J.-Y. Liu, and
S.-H. Zhu, Phys. Rev. D 78, 095006 (2008).


\end{thebibliography}
\end{document}